\documentclass[aps,prb,reprint]{revtex4-2}
\usepackage{amssymb,amsmath,bm,braket,txfonts,graphicx}
\usepackage[bookmarks=true,colorlinks,citecolor=blue,urlcolor=blue]{hyperref}

\begin{document}

\title{Effective field theory and thermal Hall effect of magnons in square-lattice antiferromagnets}
\author{Masataka Kawano}
\affiliation{Department of Basic Science, University of Tokyo, Meguro-ku, Tokyo 153-8902, Japan}
\date{\today}

\begin{abstract}
Thermal Hall transport has emerged as a powerful probe of neutral quasiparticles and associated gauge fields in insulating materials. Although the emergence of a thermal Hall effect is known to be sensitive to lattice geometry and gauge structures, an intuitive understanding of the conditions for its emergence remains limited, especially for edge-shared lattice geometries such as square and triangular lattices. Here, we develop an effective field theory of magnons in square-lattice antiferromagnets to establish the intuitive picture that elucidates the conditions for a finite thermal Hall response. By constructing an effective field theory from a spin model on the square lattice, we show that its low-energy excitations can be described by magnons with an effective SU(2) gauge field and Zeeman field that couple to magnon's pseudospins, which reflect the two-sublattice degrees of freedom in the antiferromagnets. The field strength associated with the SU(2) gauge field acts as a pseudospin-dependent magnetic field, bending the magnon's trajectories in opposite directions depending on their pseudospin. In addition, the effective Zeeman field induces an imbalance between pseudospin up and down magnons, and the combination of these two fields gives rise to the thermal Hall effect of magnons. This intuitive picture provides a systematic classification of magnetic orders in square-lattice antiferromagnets based on the presence or absence of the thermal Hall effect. We expect that our framework can be extended to various other spin models.
\end{abstract}
\maketitle

\begin{figure*}[t]
    \centering
    \includegraphics[width=175mm]{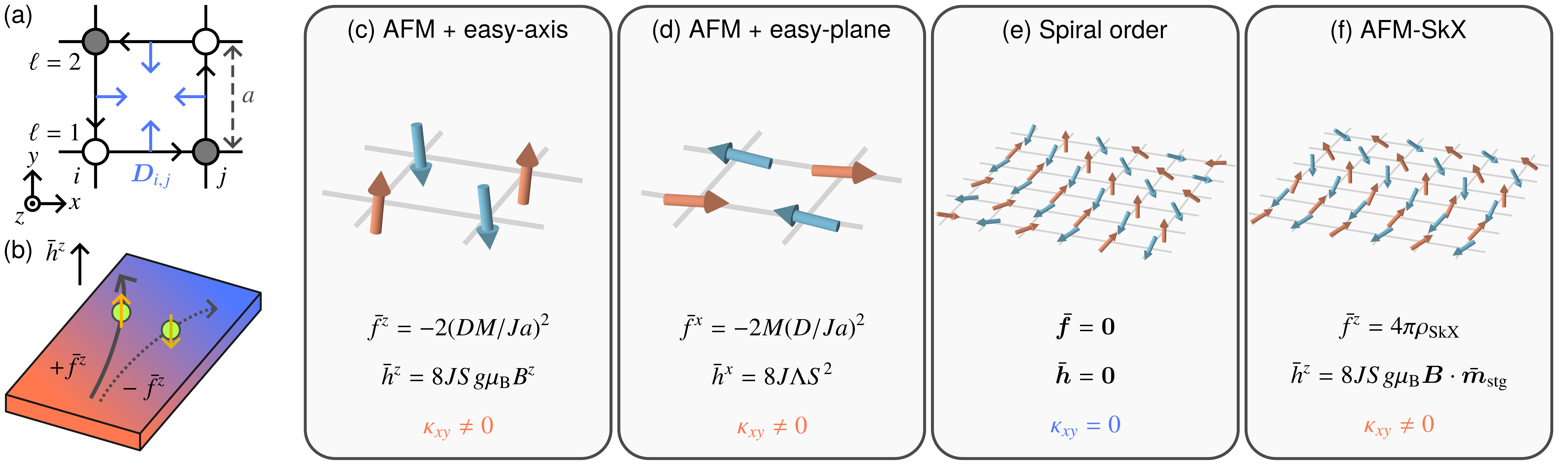}
    \caption{Model and overview of the results. (a) Schematic illustration of the spin model on the square lattice. Open and filled circles represent the two magnetic sublattices $\ell=1,2$, rendering the pseudospin degrees of freedom for magnons. The blue and black arrows denote the DM vector $\bm{D}_{i,j}=D\bm{e}^{z}\times\bm{e}_{i,j}$ and corresponding bond direction $\bm{e}_{i,j}$, respectively. The lattice constant is denoted by $a$. (b) Intuitive picture of the emergence of the thermal Hall effect from the spatially-averaged field strength $\bar{\bm{f}}$ and spatially-averaged effective Zeeman field $\bar{\bm{h}}$. The pseudospin-up and pseudospin-down magnons experience Lorentz-like force in opposite directions, and the effective Zeeman field generates the imbalance between them. The combination of these two fields leads to the finite thermal Hall conductivity $\kappa_{xy}$. (c-f) Spin configuration, corresponding spatially-averaged field strength $\bar{\bm{f}}$, spatially-averaged effective Zeeman field $\bar{\bm{h}}$, and presence or absence of $\kappa_{xy}$ for (c) easy-axis antiferromagnetic order, (d) easy-plane antiferromagnetic order, (e) spiral order with zero net magnetization $M=0$, and (f) AFM-SkX with $M\simeq0$. Orange and blue arrows represent the spin configuration on two different sublattices.}
    \label{fig:texture-gauge}
\end{figure*}

\section{Introduction}
Understanding transport properties has been a central topic in condensed matter physics as they provide key insights into low-energy excitations in material solids. Among them, the thermal Hall effect, which refers to the generation of a transverse heat current in response to a temperature gradient, has been recognized as a powerful probe for low-energy excitations in insulating materials~\cite{murakami2017jpsj,zhang2024pr}. Unlike the conventional Hall effect, which arises from the Lorentz force acting on charge carriers in an applied magnetic field, the thermal Hall effect can manifest even in insulating materials where there are no mobile charge carriers. There, the transverse heat current is carried by charge-neutral quasiparticles instead, and it emergence is driven solely by effective gauge fields, or fictitious magnetic fields, arising from microscopic interactions or symmetry-breaking patterns, thereby providing information regarding the microscopic properties of insulating materials~\cite{murakami2017jpsj,zhang2024pr}.

Magnons, charge-neutral quasiparticles corresponding to collective spin-wave excitations in magnetic insulators, can contribute to the thermal Hall effect. The thermal Hall effect in magnetic insulators was first observed in pyrochlore ferromagnet Lu$_{2}$V$_{2}$O$_{7}$~\cite{onose2010science}. Since its discovery, the thermal Hall effect has been explored in various ferromagnetic insulators: pyrochlore ferromagnets Ho$_{2}$V$_{2}$O$_{7}$ and In$_{2}$Mn$_{2}$O$_{7}$~\cite{ideue2012prb}, ferromagnetic perovskite oxide BiMnO$_{3}$~\cite{ideue2012prb}, kagome ferromagnet Cu(1-3, bdc)~\cite{hirschberger2015prl}, and ferromagnetic skyrmion crystal (FM-SkX) GaV$_{4}$Se$_{8}$~\cite{akazawa2022prr}. On the theoretical front, the intrinsic origin of the thermal Hall effect in ferromagnetic insulators can be understood by a U(1) gauge-field picture; in the presence of asymmetric exchange interactions such as Dzyaloshinskii-Moriya (DM) interaction~\cite{dzyaloshinskii1958jpcs,moriya1960pr} or noncoplanar spin texture in real space, magnons often acquire a phase factor when they move along a closed loop~\cite{fujimoto2009prl,katsura2010prl}. This phase accumulation can be interpreted as the result of magnons experiencing an effective U(1) gauge field, analogous to the phase acquired by charged particles moving in a magnetic field~\cite{fujimoto2009prl,katsura2010prl}. The associated Berry curvature in momentum space leads to a transverse deflection of magnon propagation, giving rise to the thermal Hall effect~\cite{fujimoto2009prl,katsura2010prl,matsumoto2011prb,matsumoto2011prl,matsumoto2014prb}.

While the presence of an effective U(1) gauge field is a necessary condition for the thermal Hall effect in ferromagnets, it is not always sufficient; the emergence of a finite thermal Hall response is strongly influenced by lattice geometry~\cite{katsura2010prl,ideue2012prb}. In general, DM interactions and noncoplanar spin textures yield a zero net flux, as the accumulated phase factors cancel out when summed over the entire lattice. This cancellation holds for all lattice geometries, except in cases where a uniform flux is introduced by topologically-nontrivial spin texture in real space~\cite{hoogdalem2013prb,kong2013prl,iwasaki2014prb,oh2015prb,roldan2016njp,mook2017prb,kim2019prl,nikolic2020prb,weber2022science}, Aharonov-Casher effect~\cite{nakata2017prb1,nakata2017prb2}, or orbital magnetic field~\cite{samajdar2019prb,ding2024arxiv}. In edge-shared lattices such as square and triangular lattices, the flux with opposite sign through neighboring unit cells is related by symmetry, enforcing an effective time-reversal symmetry that forbids the thermal Hall effect~\cite{katsura2010prl,ideue2012prb}. In contrast, corner-shared lattices such as kagome and pyrochlore lattices lack such symmetry constraints, allowing for a nonzero thermal Hall response even with zero net flux. This geometry constraint is not limited to ferromagnets but applies to all magnon systems described by the effective U(1) gauge-field picture. In fact, many theoretical studies on the magnon thermal Hall effect have focused exclusively on lattices with corner-shared structures: kagome~\cite{mook2014prb,lee2015prb,mook2016prb,owerre2017prb,owerre2017epl,owerre2018prb,seshadri2018prb,laurell2018prb,lu2019prb,mook2019prb}, pyrochlore~\cite{onose2010science,laurell2017prl}, honeycomb~\cite{owerre2016jap,owerre2016prb,owerre2017jap,mcclarty2018prb,neumann2022prl}, Lieb~\cite{cao2015jpcm}, star~\cite{owerre2017jpcm1,owerre2017jpcm2}, and Shastry-Sutherland lattices~\cite{judit2015ncom,malki2019prb}.

While edge-shared lattice geometries generally prohibit the thermal Hall effect due to symmetry constraints, an alternative route exists via higher-rank effective gauge fields~\cite{kawano2019prb,takeda2024ncom,kawano2024arxiv}. Unlike U(1) gauge fields, which commute and thus respect the effective time-reversal symmetry in edge-shared lattices, higher-rank gauge fields can break this symmetry due to their noncommutative nature. A well-known example in electronic systems is the anomalous Hall effect in the square lattice with Rashba spin-orbit coupling (SOC) and an external magnetic field~\cite{culcer2003prb,dugaev2005prb}, where the SOC can be interpreted as an SU(2) gauge field that couples electron spins to their kinetic motion~\cite{frohlich1993rmp}. A similar situation arises in insulating antiferromagnets, where the two-sublattice structure provides an internal degree of freedom that can be regarded as a magnonic pseudospin, analogous to the electron spin~\cite{kawano2019prb}. Although the previous study has proposed an SU(2) gauge-field picture for square-lattice antiferromagnets, a concrete formulation has remained elusive. More recently, an SU(3) gauge-field description has been developed for magnons in three-sublattice antiferromagnetic skyrmion crystals (AFM-SkXs)~\cite{takeda2024ncom,kawano2024arxiv}, highlighting the potential of higher-rank gauge fields in understanding the emergence of the thermal Hall effect in edge-shared lattices.

In this paper, we develop a low-energy effective field theory of magnons in two-sublattice antiferromagnets based on an effective SU(2) gauge-field picture. Starting from a spin model on the square lattice, we first show that the low-energy excitations are described by magnons that carry pseudospin degrees of freedom, which reflect the two-sublattice structure in the antiferromagnets. By deriving the equation of motion for magnons, we further show that magnons are coupled to an effective SU(2) gauge field and Zeeman field, depending on the spin configuration in the ground state. The field strength associated with the SU(2) gauge field acts as a pseudospin-dependent magnetic field, bending the magnon's trajectories in opposite directions for the two pseudospin components. Moreover, the effective Zeeman field induces an imbalance between these pseudospin channels, and the interplay between these two fields gives rise to the thermal Hall effect. By considering several spin configurations in the ground state, we also demonstrate how the noncommutative nature of the SU(2) gauge field circumvents the symmetry constraints that otherwise prohibit the thermal Hall effect in edge-shared lattices.

The rest of the paper is organized as follows. In Sec.~\ref{sec:model-overview}, we introduce the spin model on the square lattice and summarize our main results. In Sec.~\ref{sec:eft}, we develop the effective field theory of magnons and apply it to various ground-state spin configurations to clarify the conditions under which the thermal Hall effect emerges. Finally, in Sec.~\ref{sec:conclusion}, we conclude with a summary of our findings and an outlook for future research directions.

\section{Model and overview of the results}
\label{sec:model-overview}
Before explaining the details of the effective field theory, we first introduce the spin model that serves as the starting point of our analysis and discuss the possible magnetic orders. Then we provide an overview of our key findings: the effective field-theoretical formulation that describes the thermal Hall effect of magnons in terms of an effective SU(2) gauge field and Zeeman field. These results establish an intuitive framework for understanding the thermal Hall effect in square-lattice antiferromagnets.

\subsection{Model}
We consider a spin Hamiltonian on the square lattice,
\begin{align}
    \hat{\mathcal{H}}_{\mathrm{spin}}&=\sum_{\braket{i,j}}\left[J\hat{\bm{S}}_{i}\cdot\hat{\bm{S}}_{j}+\bm{D}_{i,j}\cdot(\hat{\bm{S}}_{i}\times\hat{\bm{S}}_{j})\right]\nonumber\\
    &\quad+\sum_{i}\left[\Lambda(\hat{S}_{i}^{z})^{2}-g\mu_{\mathrm{B}}\bm{B}\cdot\hat{\bm{S}}_{i}\right],
    \label{eq:Hspin}
\end{align}
where $\sum_{\braket{i,j}}$ runs over all pairs of nearest-neighbor $i$ and $j$ sites, $\hat{\bm{S}}_{i}=(\hat{S}_{i}^{x},\hat{S}_{i}^{y},\hat{S}_{i}^{z})$ is the spin-$S$ operator at site $i$, $J$ is the Heisenberg exchange coupling, $\bm{D}_{i,j}$ is the DM interaction, $\Lambda$ is the single-ion anisotropy, and $\bm{B}$ is the magnetic field with g-factor $g$ and Bohr magneton $\mu_{\mathrm{B}}$. The alignment of the DM vector is determined by the lattice symmetry; we assume the $C_{4v}$ point group symmetry, which yields $\bm{D}_{i,j}=D\bm{e}^{z}\times\bm{e}_{i,j}$, where $D=|\bm{D}_{i,j}|$, $\bm{e}^{\alpha}$ ($\alpha=x,y,z$) is the unit vector pointing in $\alpha$-direction in spin space, and $\bm{e}_{i,j}$ is the unit vector pointing from site $i$ to nearest-neighbor site $j$. The schematic illustration of the model is shown in Fig.~\ref{fig:texture-gauge}(a).

Various spin configurations are realized in the ground state by the competition between the parameters in the Hamiltonian~(\ref{eq:Hspin})~\cite{bogdanov1989jetp,bogdanov2002prb}. In the absence of the single-ion anisotropy and magnetic field, the system favors a two-fold degenerate spiral order, with the pitch determined by the ratio $|D/J|$. A sufficiently large single-ion anisotropy stabilizes the canted antiferromagnetic order. A numerical study also indicates that when a sufficiently strong magnetic field is applied with the moderate single-ion anisotropy, the linear combination of two-degenerate spiral states becomes energetically favorable, stabilizing the AFM-SkX phase~\cite{keesman2015prb}.

The model Hamiltonian (\ref{eq:Hspin}) with $\Lambda\geq0$ and $\pi$-rotation about the $x=y$ axis can serve as the effective spin model for noncentrosymmetric magnetic insulators Ba$_{2}$XGe$_{2}$O$_{7}$ ($X=$ Cu, Co, Mn), where the $X$O$_{4}$ tetrahedra form a square lattice~\cite{zheludev1996prb,zheludev1997prl,zheludev1997prb,zheludev1999prb,sato2003pbcm,zheludev2003prb,yi2008apl,masuda2010prb,murakawa2012prb}. The canted antiferromagnetic order is observed below $4.0$K in Ba$_{2}$MnGe$_{2}$O$_{7}$~\cite{masuda2010prb} and below $6.7$K in Ba$_{2}$CoGe$_{2}$O$_{7}$~\cite{zheludev2003prb}. In Ba$_{2}$CuGe$_{2}$O$_{7}$, Cu$^{2+}$ ion has spin-$1/2$ and single-ion magnetic anisotropy is absent, leading to the spiral order below $3.0$K~\cite{zheludev1996prb}. The model (\ref{eq:Hspin}) with $S=1/2$ can also serve as the effective spin model for K$_{2}$V$_{3}$O$_{8}$~\cite{lumsden2001prl}.

\subsection{Overview of the results}
Here, we briefly summarize our findings. The key step in our formulation is to identify the two-sublattice structure in antiferromagnets as pseudospin degrees of freedom for magnons. As a result, the low-energy excitations can be described by two species of magnons, represented by bosonic operators $\hat{b}_{\ell}(\bm{r})$, where $\ell=1,2$ indexes the pseudospin and $\bm{r}$ is the spatial position. Our main result is that, by deriving the low-energy effective field theory, the equation of motion for the two-component field $\hat{\bm{b}}(\bm{r},t)=(\hat{b}_{1}(\bm{r},t),\hat{b}_{2}(\bm{r},t))^{T}$ takes the following form
\begin{align}
    &-\hbar^{2}\partial_{t}^{2}\hat{\bm{b}}(\bm{r},t)\nonumber\\
    &=\left[-8(JSa)^{2}\left(\bm{\nabla}\sigma^{0}-i\bm{T}(\bm{r})\right)^{2}+u(\bm{r})\sigma^{0}+\bm{h}(\bm{r})\cdot\bm{\sigma}\right]\hat{\bm{b}}(\bm{r},t).
    \label{eq:eom-b}
\end{align}
Here, $\bm{T}(\bm{r})=(T_{x}(\bm{r}),T_{y}(\bm{r}))$ is a set of $2\times2$ Hermitian and traceless matrices, $u(\bm{r})$ is a potential term, and $\bm{h}(\bm{r})=(h^{x}(\bm{r}),h^{y}(\bm{r}),h^{z}(\bm{r}))$ describes an effective Zeeman field that couples to the pseudospins. The matrix $T(\bm{r})$ belongs to the SU(2) Lie algebra and can be regarded as an effective SU(2) gauge field, describing the coupling between the pseudospin and the kinetic motion of magnons, analogous to how a U(1) gauge field, or vector potential, couples to charged particles.

From $\bm{T}(\bm{r})$, we define a field strength
\begin{align}
    F(\bm{r})=\partial_{x}T_{y}(\bm{r})-\partial_{y}T_{x}(\bm{r})-i[T_{x}(\bm{r}),T_{y}(\bm{r})],
    \label{eq:F}
\end{align}
which consists of two contributions: the first two terms resemble the conventional U(1) magnetic field associated with a vector potential, while the last term, $-i[T_{x}(\bm{r}),T_{y}(\bm{r})]$, reflects the noncommutative nature of the SU(2) gauge field. Since $F(\bm{r})$ also belongs to the SU(2) Lie algebra, it can be expressed as
\begin{align}
    F(\bm{r})=\bm{f}(\bm{r})\cdot\bm{\sigma},
    \label{eq:F=fsigma}
\end{align}
and $\bm{f}(\bm{r})$ can be interpreted as a pseudospin-dependent magnetic field~\cite{hatano2007pra}, which bends the propagation of magnons with different pseudospins in opposite transverse directions.

In the low-energy and long-wavelength limit, the uniform part of the field strength and effective Zeeman field is expected to be dominant, and we only focus on the spatially-averaged fields, $\bar{\bm{f}}=(1/V)\int\mathrm{d}^{2}\bm{r}\ \bm{f}(\bm{r})$ and $\bar{\bm{h}}=(1/V)\int\mathrm{d}^{2}\bm{r}\ \bm{h}(\bm{r})$, where $V$ is the volume of the system. The intuitive picture for the case $\bar{\bm{f}}=(0,0,\bar{f}^{z})$ and $\bar{\bm{h}}=(0,0,\bar{h}^{z})$ is illustrated in Fig.~\ref{fig:texture-gauge}(b). In this scenario, pseudospin-up and down magnons experience opposite Lorentz-like forces, leading to their separation in real space. For a finite thermal Hall effect, an imbalance between the contributions of the two pseudospin magnons is required, which is induced by the effective Zeeman field $\bar{h}^{z}$.

To demonstrate the utility of this intuitive framework, we apply the SU(2) gauge field description to four distinct spin textures that can be realized in the spin Hamiltonian~(\ref{eq:Hspin}). Figures~\ref{fig:texture-gauge}(c)-\ref{fig:texture-gauge}(f) illustrate the spin configurations, spatially-avaraged field strengths $\bar{\bm{f}}$, spatially-averaged effective Zeeman field $\bar{\bm{h}}$, and presence or absence of the thermal Hall conductivity $\kappa_{xy}$ for (c) easy-axis antiferromagnetic order, (d) easy-plane antiferromagnetic order, (e) spiral order with zero net magnetization $M=0$, and (f) AFM-SkX with $M\simeq0$. In both easy-axis and easy-plane antiferromagnetic orders shown in Figs.~\ref{fig:texture-gauge}(c) and~\ref{fig:texture-gauge}(d), we find nonzero field strength $\bar{\bm{f}}$ and effective Zeeman field $\bar{\bm{h}}$ that are parallel to each other, leading to a finite thermal Hall conductivity $\kappa_{xy} \neq 0$. The two effective fields arise from a combination of net magnetization $M$, DM interaction $D$, and either an external magnetic field $\bm{B}$ or single-ion anisotropy $\Lambda$. By contrast, for the spiral order shown in Fig.~\ref{fig:texture-gauge}(e), the field strength and effective Zeeman field vanish on average, enforcing $\kappa_{xy} = 0$. Finally, in the AFM-SkX phase shown in Fig.\ref{fig:texture-gauge}(f), the average field strength is generated by the skyrmion density $\rho_{\mathrm{SkX}}$ defined as Eq.~(\ref{eq:skx}), and the combination of the effective Zeeman field results in a finite $\kappa_{xy}$. This analysis clarifies how exchange interactions, anisotropy, external magnetic field, and real-space spin textures contribute to the thermal Hall effect.

\section{Effective field theory of magnons}
\label{sec:eft}
In this section, we develop the effective field theory of magnons in square-lattice antiferromagnets. Starting with the spin Hamiltonian (\ref{eq:Hspin}), we construct a low-energy effective field theory that captures the essential physics of magnons in the antiferromagnets. From the equation of motion for magnons, we further show that the system can be effectively described by magnons coupled to an SU(2) gauge field that encodes the effect of the DM interaction and real-space spin textures. Then we demonstrate that the SU(2) gauge-field picture clarifies how various terms in the Hamiltonian and real-space spin textures break the time-reversal symmetry and contribute to the thermal Hall effect of magnons.

\subsection{Effective magnon Hamiltonian}
To describe the low-energy and long-wavelength excitations, we begin by taking the continuum limit, where the system is expected to be described by two slowly varying spin density operators $\hat{\bm{s}}_{\ell}(\bm{r})$ ($\ell=1,2$), incorporating the two-sublattice structure of the antiferromagnets. We also introduce two slowly varying unit vectors $\bm{m}_{\ell}(\bm{r})$ that represent the ground-state spin configuration on the sublattice $\ell$. The effective field theory can be constructed from Eq.~(\ref{eq:Hspin}) by replacing the discrete sums and spin operators with their continuous counterparts: $\sum_{i}\to(1/v)\int\mathrm{d}^{2}\bm{r}\sum_{\ell}$ and $\hat{\bm{S}}_{i}\to v\hat{\bm{s}}_{\ell}(\bm{r})$ for site $i$ that belongs to the sublattice $\ell$, where $v=2a^{2}$ is the volume per unit cell and $a$ is the lattice constant. Including terms up to the second-order derivatives of $\hat{\bm{s}}_{\ell}(\bm{r})$, the effective spin Hamiltonian is expressed as
\begin{align}
    \hat{\mathcal{H}}_{\mathrm{eff}}&=\int\mathrm{d}^{2}\bm{r}\left[\sum_{\ell,\ell'}\hat{\bm{s}}_{\ell}^{T}(\bm{r})K_{\ell,\ell'}(\bm{r})\hat{\bm{s}}_{\ell'}(\bm{r})-g\mu_{\mathrm{B}}\bm{B}\cdot\sum_{\ell}\hat{\bm{s}}_{\ell}(\bm{r})\right].
\end{align}
The $3\times3$ matrix $K_{\ell,\ell'}(\bm{r})$ encodes the effects of the exchange interactions and single-ion anisotropy,
\begin{align}
    K_{1,1}(\bm{r})&=K_{2,2}(\bm{r})=\Lambda v
    \begin{pmatrix}
        0 & 0 & 0\\
        0 & 0 & 0\\
        0 & 0 & 1
    \end{pmatrix}
    ,\\
    K_{1.2}(\bm{r})&=K_{2.1}(\bm{r})=2Jv\left(1+\frac{a^{2}}{4}\nabla^{2}\right)I+Dva
    \begin{pmatrix}
        0 & 0 & -\partial_{x} \\
        0 & 0 & -\partial_{y} \\
        \partial_{x} & \partial_{y} & 0
    \end{pmatrix}
    ,
\end{align}
where $I$ is the $3\times3$ identity matrix.

The transformation from spin systems to magnon systems is performed by the Holstein-Primakoff transformation~\cite{holstein1940pr}
\begin{align}
    \hat{\bm{s}}_{\ell}(\bm{r})\simeq\sqrt{\frac{S}{v}}\left(\hat{b}_{\ell}(\bm{r})\bm{e}_{\ell}^{-}(\bm{r})+\mathrm{H.c.}\right)+\left(\frac{S}{v}-\hat{b}_{\ell}^{\dagger}(\bm{r})\hat{b}_{\ell}(\bm{r})\right)\bm{m}_{\ell}(\bm{r}),
\end{align}
where $\bm{e}_{\ell}^{\pm}(\bm{r})=(1/\sqrt{2})(\bm{e}_{\ell}^{X}(\bm{r})\pm i\bm{e}_{\ell}^{Y}(\bm{r}))$ and the three unit vectors, $\bm{e}_{\ell}^{X}(\bm{r})$, $\bm{e}_{\ell}^{Y}(\bm{r})$, and $\bm{m}_{\ell}(\bm{r})$, form an orthonormal basis. To carry out the calculations, the directions of $\bm{e}_{\ell}^{X}(\bm{r})$ and $\bm{e}_{\ell}^{Y}(\bm{r})$ must be specified, corresponding to the gauge fixing. For this purpose, we introduce the normalized uniform and staggered magnetization vectors, $\bm{m}_{\mathrm{uni}}(\bm{r})$ and $\bm{m}_{\mathrm{stg}}(\bm{r})$, as
\begin{align}
    \bm{m}_{1}(\bm{r})+\bm{m}_{2}(\bm{r})&=2M\bm{m}_{\mathrm{uni}}(\bm{r}),\\
    \bm{m}_{1}(\bm{r})-\bm{m}_{2}(\bm{r})&=2\sqrt{1-M^{2}}\bm{m}_{\mathrm{stg}}(\bm{r}),
\end{align}
where $M$ is the amplitude of the uniform magnetization. Here, we assume that $M$ is spatially uniform and $M/J\ll1$, allowing us to neglect second and higher-order terms in $M$. The unit vector $\bm{e}_{\ell}^{X}(\bm{r})$ is choosen as $\bm{e}_{1}^{X}(\bm{r})=\bm{e}_{2}^{X}(\bm{r})=\bm{e}^{X}(\bm{r})=\bm{m}_{\mathrm{uni}}(\bm{r})\times\bm{m}_{\mathrm{stg}}(\bm{r})$. The remaining unit vector $\bm{e}_{\ell}^{Y}(\bm{r})$ is determined automatically by the orthogonality condition. We neglect high-energy contributions, such as magnon-magnon interactions~\cite{zhitomirsky2013rmp} and terms quadratic in the derivatives of $\bm{e}_{\ell}^{\pm}(\bm{r})$. The constant term is also ignored as it does not influence the following discussion. The effective magnon Hamiltonian is then obtained as
\begin{align}
    \hat{\mathcal{H}}_{\mathrm{eff}}\simeq\frac{1}{2}\int\mathrm{d}^{2}\bm{r}\ \hat{\Phi}^{\dagger}(\bm{r})H(\bm{r})\hat{\Phi}(\bm{r}),
    \label{eq:Heff}
\end{align}
where $\hat{\Phi}(\bm{r})=(\hat{b}_{1}(\bm{r}),\hat{b}_{1}^{\dagger}(\bm{r}),\hat{b}_{2}(\bm{r}),\hat{b}_{2}^{\dagger}(\bm{r}))^{T}$ and $H(\bm{r})$ is the $4\times4$ matrix defined as
\begin{align}
    H(\bm{r})=4JS(\sigma^{0}\otimes\tau^{0}+\sigma^{x}\otimes\tau^{x})+\sum_{\mu,\nu=0,x,y,z}g_{\mu,\nu}(\bm{r})\sigma^{\mu}\otimes\tau^{\nu},
    \label{eq:H(r)}
\end{align}
where
\begin{align}
    g_{0,0}(\bm{r})&=\Lambda S\left[(\bm{e}^{X}(\bm{r})\cdot\bm{e}^{z})^{2}+(m_{\mathrm{uni}}^{z}(\bm{r}))^{2}-2(m_{\mathrm{stg}}^{z}(\bm{r}))^{2}\right]\nonumber\\
    &\quad+g\mu_{\mathrm{B}}M\bm{B}\cdot\bm{m}_{\mathrm{uni}}(\bm{r}),\\
    g_{y,0}(\bm{r})&=-2iJMSa^{2}[(\bm{m}_{\mathrm{uni}}(\bm{r}))^{T}\bm{\nabla}\bm{m}_{\mathrm{stg}}(\bm{r})]\cdot\bm{\nabla}\nonumber\\
    &\quad-2iDMSa\left[(\bm{e}^{X}(\bm{r})\cdot\bm{e}^{y})\partial_{x}-(\bm{e}^{X}(\bm{r})\cdot\bm{e}^{x})\partial_{y}\right],\\
    g_{z,0}(\bm{r})&=-6\Lambda MSm_{\mathrm{uni}}^{z}(\bm{r})m_{\mathrm{stg}}^{z}(\bm{r})+g\mu_{\mathrm{B}}\bm{B}\cdot\bm{m}_{\mathrm{stg}}(\bm{r}),\\
    g_{0,x}(\bm{r})&=\Lambda S\left[(\bm{e}^{X}(\bm{r})\cdot\bm{e}^{z})^{2}-(m_{\mathrm{uni}}^{z}(\bm{r}))^{2}\right],\\
    g_{x,x}(\bm{r})&=JSa^{2}\nabla^{2},\\
    g_{y,x}(\bm{r})&=2iJMSa^{2}[(\bm{m}_{\mathrm{uni}}(\bm{r}))^{T}\bm{\nabla}\bm{m}_{\mathrm{stg}}(\bm{r})]\cdot\bm{\nabla}\nonumber\\
    &\quad+2iDMSa\left[(\bm{e}^{X}(\bm{r})\cdot\bm{e}^{y})\partial_{x}-(\bm{e}^{X}(\bm{r})\cdot\bm{e}^{x})\partial_{y}\right],\\
    g_{z,x}(\bm{r})&=2\Lambda MSm_{\mathrm{uni}}^{z}(\bm{r})m_{\mathrm{stg}}^{z}(\bm{r}),\\
    g_{0,y}(\bm{r})&=\Lambda MS(\bm{e}^{X}(\bm{r})\cdot\bm{e}^{z})m_{\mathrm{stg}}^{z}(\bm{r}),\\
    g_{y,y}(\bm{r})&=2iJSa^{2}\bm{A}_{\mathrm{stg}}(\bm{r})\cdot\bm{\nabla}\nonumber\\
    &\quad-2iDSa(m_{\mathrm{stg}}^{y}(\bm{r})\partial_{x}-m_{\mathrm{stg}}^{x}(\bm{r})\partial_{y}),\\
    g_{z,y}(\bm{r})&=-\Lambda S(\bm{e}^{X}(\bm{r})\cdot\bm{e}^{z})m_{\mathrm{uni}}^{z}(\bm{r}),\\
    g_{x,z}(\bm{r})&=-2iJSa^{2}\bm{A}_{\mathrm{uni}}(\bm{r})\cdot\bm{\nabla}\nonumber\\
    &\quad-2iDMSa(m_{\mathrm{uni}}^{y}(\bm{r})\partial_{x}-m_{\mathrm{uni}}^{x}(\bm{r})\partial_{y}),
\end{align}
and $g_{\mu,\nu}(\bm{r})=0$ for the other $\mu$ and $\nu$. Here, we define unit and Pauli matrices $\sigma^{\mu}$ and $\tau^{\mu}$ ($\mu=0,x,y,z$) representing the sublattice and particle-hole degrees of freedom, respectively. The two-component real vector fields $\bm{A}_{\mathrm{uni}}(\bm{r})=(A_{\mathrm{uni},x}(\bm{r}),A_{\mathrm{uni},y}(\bm{r}))$ and $\bm{A}_{\mathrm{stg}}(\bm{r})=(A_{\mathrm{stg},x}(\bm{r}),A_{\mathrm{stg},y}(\bm{r}))$ are defined as
\begin{align}
    \bm{A}_{\mathrm{uni}}(\bm{r})&=\frac{i}{2}\left([\bm{e}_{1}^{+}(\bm{r})]^{T}\bm{\nabla}\bm{e}_{1}^{-}(\bm{r})+[\bm{e}_{2}^{+}(\bm{r})]^{T}\bm{\nabla}\bm{e}_{2}^{-}(\bm{r})\right),\\
    \bm{A}_{\mathrm{stg}}(\bm{r})&=\frac{i}{2}\left([\bm{e}_{1}^{+}(\bm{r})]^{T}\bm{\nabla}\bm{e}_{1}^{-}(\bm{r})-[\bm{e}_{2}^{+}(\bm{r})]^{T}\bm{\nabla}\bm{e}_{2}^{-}(\bm{r})\right),
\end{align}
which can be regarded as the linear combination of the vector potential $i[\bm{e}_{\ell}^{+}(\bm{r})]^{T}\bm{\nabla}\bm{e}_{\ell}^{-}(\bm{r})$ that arises from the real-space spin texture on the sublattice $\ell$.

\subsection{Equation of motion and SU(2) gauge field}
The gauge structure of the effective magnon Hamiltonian in Eq.~(\ref{eq:Heff}) is obscured by its complexity. To clarify this structure, we derive the equation of motion for magnons. The Heisenberg equation of motion for $\hat{\Phi}(\bm{r},t)$ is given by
\begin{align}
    i\hbar\partial_{t}\hat{\Phi}(\bm{r},t)=(\sigma^{0}\otimes\tau^{z})H(\bm{r})\hat{\Phi}(\bm{r},t).
    \label{eq:eom-Phi-first}
\end{align}
Since the equation of motion for magnons in antiferromagnets is expected to be described with second-order time derivatives, we use Eq.~(\ref{eq:eom-Phi-first}) twice to have
\begin{align}
    -\hbar^{2}\partial_{t}^{2}\hat{\Phi}(\bm{r},t)=(\sigma^{0}\otimes\tau^{z})H(\bm{r})(\sigma^{0}\otimes\tau^{z})H(\bm{r})\hat{\Phi}(\bm{r},t).
    \label{eq:eom-Phi-second}
\end{align}
For simplicity, we assume that the Heisenberg exchange is dominant, namely $|D/J|$, $|\Lambda/J|$, $|g\mu_{\mathrm{B}}\bm{B}/J|\ll1$, which leads to  $g_{\mu,\nu}(\bm{r})/4JS\ll1$. Then the matrix in the right-hand side of Eq.~(\ref{eq:eom-Phi-second}) can be approximated as
\begin{align}
    &(\sigma^{0}\otimes\tau^{z})H(\bm{r})(\sigma^{0}\otimes\tau^{z})H(\bm{r})\nonumber\\
    &\simeq8JS\left(\sum_{\mu}g_{\mu,0}(\bm{r})\sigma^{\mu}-g_{0,x}(\bm{r})\sigma^{x}-g_{x,x}(\bm{r})\sigma^{0}\right)\otimes\tau^{0}\nonumber\\
    &\quad+8JS\left(g_{y,y}(\bm{r})\sigma^{z}-g_{z,y}(\bm{r})\sigma^{y}+\sum_{\mu}g_{\mu,z}(\bm{r})\sigma^{\mu}\right)\otimes\tau^{z}.
    \label{eq:tauzHtauzH}
\end{align}
We remark that off-diagonal terms in the particle-hole space, such as $g_{y,0}(\bm{r})\sigma^{z}\otimes\tau^{x}$, only give higher-order contributions and can be neglected. The matrix in Eq.~(\ref{eq:tauzHtauzH}) now has the diagonal form in the particle-hole space, and the equation of motion (\ref{eq:eom-Phi-second}) can be reduced to Eq.~(\ref{eq:eom-b}) with
\begin{align}
    T_{x}(\bm{r})&=\left(-A_{\mathrm{uni},x}(\bm{r})-\frac{DM}{Ja}m_{\mathrm{uni}}^{y}(\bm{r})\right)\sigma^{x}\nonumber\\
    &\quad+\left(-M\bm{m}_{\mathrm{uni}}(\bm{r})\cdot\partial_{x}\bm{m}_{\mathrm{stg}}(\bm{r})-\frac{DM}{Ja}\bm{e}^{X}(\bm{r})\cdot\bm{e}^{y}\right)\sigma^{y}\nonumber\\
    &\quad+\left(A_{\mathrm{stg},x}(\bm{r})-\frac{D}{Ja}m_{\mathrm{stg}}^{y}(\bm{r})\right)\sigma^{z},
    \label{eq:Tx}\\
    T_{y}(\bm{r})&=\left(-A_{\mathrm{uni},y}(\bm{r})+\frac{DM}{Ja}m_{\mathrm{uni}}^{x}(\bm{r})\right)\sigma^{x}\nonumber\\
    &\quad+\left(-M\bm{m}_{\mathrm{uni}}(\bm{r})\cdot\partial_{x}\bm{m}_{\mathrm{stg}}(\bm{r})+\frac{DM}{Ja}\bm{e}^{X}(\bm{r})\cdot\bm{e}^{x}\right)\sigma^{y}\nonumber\\
    &\quad+\left(A_{\mathrm{stg},y}(\bm{r})+\frac{D}{Ja}m_{\mathrm{stg}}^{x}(\bm{r})\right)\sigma^{z},
    \label{eq:Ty}\\
    u(\bm{r})&=8J\Lambda S^{2}\left[(\bm{e}^{X}(\bm{r})\cdot\bm{e}^{z})^{2}+(m_{\mathrm{uni}}^{z}(\bm{r}))^{2}-2(m_{\mathrm{stg}}^{z}(\bm{r}))^{2}\right]\nonumber\\
    &\quad+8JSg\mu_{\mathrm{B}}M\bm{B}\cdot\bm{m}_{\mathrm{uni}}(\bm{r}),\\
    h^{x}(\bm{r})&=-8J\Lambda S^{2}\left[(\bm{e}^{X}(\bm{r})\cdot\bm{e}^{z})^{2}-(m_{\mathrm{uni}}^{z}(\bm{r}))^{2}\right],\\
    h^{y}(\bm{r})&=8J\Lambda S^{2}(\bm{e}^{X}(\bm{r})\cdot\bm{e}^{z})m_{\mathrm{uni}}^{z}(\bm{r}),\\
    h^{z}(\bm{r})&=-48J\Lambda MS^{2}m_{\mathrm{uni}}^{z}(\bm{r})m_{\mathrm{stg}}^{z}(\bm{r})+8JSg\mu_{\mathrm{B}}\bm{B}\cdot\bm{m}_{\mathrm{stg}}(\bm{r}).
\end{align}
For later convenience, we introduce the three-component real vector field $\bm{t}_{\alpha}(\bm{r})=(t_{\alpha}^{x}(\bm{r}),t_{\alpha}^{y}(\bm{r}),t_{\alpha}^{z}(\bm{r}))$ ($\alpha=x,y$) as
\begin{align}
    T_{\alpha}(\bm{r})=\bm{t}_{\alpha}(\bm{r})\cdot\bm{\sigma}.
    \label{eq:T=tsigma}
\end{align}
Without the SU(2) gauge field, potential term, and effective Zeeman field term, the equation of motion (\ref{eq:eom-b}) reproduces the well-known two-degenerate magnon bands in two-dimensional antiferromagnets, $\varepsilon(\bm{k})\simeq2\sqrt{2}(JSa/\hbar)|\bm{k}|$, where $\bm{k}$ is the wave vector. The SU(2) gauge field $\bm{T}(\bm{r})$ generates a physical field, called field strength $F$, defined as Eq.~(\ref{eq:F}). The three-component real vector field $\bm{f}(\bm{r})$ in Eq.~(\ref{eq:F=fsigma}) is calculated as
\begin{align}
    \bm{f}(\bm{r})=\partial_{x}\bm{t}_{y}(\bm{r})-\partial_{y}\bm{t}_{x}(\bm{r})+2\bm{t}_{x}(\bm{r})\times\bm{t}_{y}(\bm{r}).
    \label{eq:f}
\end{align}

\subsection{Effective time-reversal symmetry breaking}
To have a finite thermal Hall conductivity, the equation of motion (\ref{eq:eom-b}) should break the effective time-reversal symmetry described by antiunitary operators $\hat{\mathcal{T}}_{\mathrm{I}}$ (orthogonal) and $\hat{\mathcal{T}}_{\mathrm{II}}$ (symplectic) satisfying
\begin{align}
    \hat{\mathcal{T}}_{\mathrm{I}}\hat{\bm{b}}(\bm{r},t)\hat{\mathcal{T}}_{\mathrm{I}}^{-1}&=\hat{\bm{b}}(\bm{r},-t),
    \label{eq:T_I}\\
    \hat{\mathcal{T}}_{\mathrm{II}}\hat{\bm{b}}(\bm{r},t)\hat{\mathcal{T}}_{\mathrm{II}}^{-1}&=-i\sigma^{y}\hat{\bm{b}}(\bm{r},-t).
    \label{eq:T_II}
\end{align}
The time-reversal operator $\hat{\mathcal{T}}_{\mathrm{II}}$ includes the pseudospin flip~\cite{kondo2019prb1,kondo2019prb2}. When the equation of motion (\ref{eq:eom-b}) is invariant under the time-reversal operations described by Eqs.~(\ref{eq:T_I}) or~(\ref{eq:T_II}), the system does not exhibit the finite thermal Hall conductivity~\cite{mook2019prb}. The time-reversal symmetry by $\hat{\mathcal{T}}_{\mathrm{I}}$ requires $t_{\alpha}^{x}(\bm{r})=t_{\alpha}^{z}(\bm{r})=0$ ($\alpha=x,y$) and $h^{y}(\bm{r})=0$, while the time-reversal symmetry by $\hat{\mathcal{T}}_{\mathrm{II}}$ requires $\bm{h}(\bm{r})=\bm{0}$. Therefore, to break both symmetries, finite $t_{\alpha}^{x}(\bm{r})$ or $t_{\alpha}^{z}(\bm{r})$ with $\bm{h}(\bm{r})\neq\bm{0}$ is necessary.

There are cases, however, that the seemingly time-reversal broken system is reduced to the time-reversal invariant one by applying a local SU(2) gauge transformation, which is described by a unitary operator $W_{\bm{\theta}}(\bm{r})=\exp[-i(\bm{\theta}(\bm{r})/2)\cdot\bm{\sigma}]$ with three-dimensional real vector field $\bm{\theta}(\bm{r})$. The local SU(2) gauge transformation corresponds to the reconstruction of local bosonic operators from the linear combination of $\hat{b}_{\ell}(\bm{r})$. Applying the SU(2) gauge transformation, $\hat{\bm{b}}(\bm{r},t)\to W_{\bm{\theta}}(\bm{r})\hat{\bm{b}}(\bm{r},t)$, the SU(2) gauge field $\bm{t}_{\alpha}(\bm{r})$, field strength $\bm{f}(\bm{r})$, and effective Zeeman field $\bm{h}(\bm{r})$ are transformed as
\begin{align}
    \bm{t}_{\alpha}(\bm{r})&\to R_{\bm{\theta}}^{\dagger}(\bm{r})\bm{t}_{\alpha}(\bm{r})R_{\bm{\theta}}(\bm{r})+\partial_{\alpha}\bm{\theta}(\bm{r}),\\
    \bm{f}(\bm{r})&\to R_{\bm{\theta}}^{\dagger}(\bm{r})\bm{f}(\bm{r})R_{\bm{\theta}}(\bm{r}),\\
    \bm{h}(\bm{r})&\to R_{\bm{\theta}}^{\dagger}(\bm{r})\bm{h}(\bm{r})R_{\bm{\theta}}(\bm{r}),
\end{align}
where $R_{\bm{\theta}}(\bm{r})$ is the $3\times3$ orthogonal matrix describing the three-dimensional rotation by $|\bm{\theta}|$ about $\bm{\theta}/|\bm{\theta}|$ axis. For example, when the system does not have the time-reversal symmetry but $\bm{t}_{\alpha}(\bm{r})$, $\bm{f}(\bm{r})$, and $\bm{h}(\bm{r})$ take the form as
\begin{align}
    \bm{t}_{x}(\bm{r})&=(t_{x}^{x}(\bm{r}),0,0),\\
    \bm{t}_{y}(\bm{r})&=(t_{y}^{x}(\bm{r}),0,0),\\
    \bm{f}(\bm{r})&=(\partial_{x}t_{y}^{x}(\bm{r})-\partial_{y}t_{x}^{x}(\bm{r}),0,0),\\
    \bm{h}(\bm{r})&=(0,0,h^{z}(\bm{r})),
\end{align}
with nonzero $t_{x}^{x}(\bm{r})$, $t_{y}^{x}(\bm{r})$, and $h^{z}(\bm{r})$, the SU(2) gauge transformation with $\bm{\theta}(\bm{r})=(0,0,\pi/2)$ leads to
\begin{align}
    \bm{t}_{x}(\bm{r})&=(0,t_{x}^{x}(\bm{r}),0),\\
    \bm{t}_{y}(\bm{r})&=(0,t_{y}^{x}(\bm{r}),0),\\
    \bm{f}(\bm{r})&=(0,\partial_{x}t_{y}^{x}(\bm{r})-\partial_{y}t_{x}^{x}(\bm{r}),0),\\
    \bm{h}(\bm{r})&=(0,0,h^{z}(\bm{r})).
\end{align}
The transformed system has the time-reversal symmetry described by $\hat{\mathcal{T}}_{\mathrm{I}}$. It can be intuitively understood as follows; the system has the field strength $\bm{f}(\bm{r})$ which is perpendicular to $\bm{h}(\bm{r})$, and their combined effect does not yield the required net transverse motion of magnons as shown in Fig.~\ref{fig:texture-gauge}(b). In other words, a finite thermal Hall effect generally requires that the pseudospin-dependent magnetic field $\bm{f}(\bm{r})$ and the effective Zeeman field $\bm{h}(\bm{r})$ act in the same direction, namely $\bm{f}(\bm{r})\cdot\bm{h}(\bm{r})\neq0$. In this case, there are no local SU(2) gauge transformations to recover the time-reversal symmetry. Since the spatially-averaged fields $\bar{\bm{f}}$ and $\bar{\bm{h}}$ are expected to be dominant in the low-energy and long-wavelength limit, the condition to have a finite thermal Hall conductivity can be written as
\begin{align}
    \bar{\bm{f}}\cdot\bar{\bm{h}}\neq0.
\end{align}

We briefly note the difference between the effective SU(2) and SU(3) gauge fields discussed in Refs.~\cite{takeda2024ncom,kawano2024arxiv}. For the SU(3) gauge field, the associated field strength can be characterized by an eight-dimensional real vector field $\bm{f}_{\mathrm{SU(3)}}(\bm{r})\in\mathbb{R}^{8}$. In contrast to the SU(2) case, there is generally no time-reversal symmetry operation that flips the field strength as $\bm{f}_{\mathrm{SU(3)}}(\bm{r})\to-\bm{f}_{\mathrm{SU(3)}}(\bm{r})$. Therefore, in systems described by the SU(3) gauge field, it is often sufficient to break the time-reversal symmetry represented by $\hat{\mathcal{T}}_{\mathrm{I}}$, and there might be a chance to break the time-reversal symmetry without the effective Zeeman field term.

\subsection{Applications}
Here, we apply our framework to several different magnetic orders that can be realized in our spin model.

\subsubsection{Canted antiferromagnetic order with easy-axis anisotropy}
We first consider the case where the ground-state spin configuration is given by the canted antiferromagnetic order represented by $\bm{m}_{\mathrm{uni}}(\bm{r})=\bm{e}^{x}$ and $\bm{m}_{\mathrm{stg}}(\bm{r})=\bm{e}^{z}$ as shown in Fig.~\ref{fig:texture-gauge}(c), which can be realized when $\Lambda<0$, $\bm{B}=(B^{x},0,B^{z})$, and $|\Lambda|$ is sufficiently larger than $|D|$ and $|B^{z}|$. The nonzero elements of the SU(2) gauge field, on-site potential, and effective Zeeman field are calculated From Eqs.~(\ref{eq:Tx})-(\ref{eq:T=tsigma}) as
\begin{align}
    t_{x}^{y}(\bm{r})&=\frac{DM}{Ja},\\
    t_{y}^{x}(\bm{r})&=\frac{DM}{Ja},\\
    u(\bm{r})&=-16J\Lambda S^{2}+8JSg\mu_{\mathrm{B}}MB^{x},\\
    h^{z}(\bm{r})&=8JSg\mu_{\mathrm{B}}B^{z},
\end{align}
This situation is similar to the electronic systems with Rashba SOC and magnetic field in $z$-direction. The nonzero element of the field strength comes from the commutation relation of the SU(2) gauge field $-i[T_{x}(\bm{r}),T_{y}(\bm{r})]$, and is calculated from Eq.~(\ref{eq:f}) as
\begin{align}
    f^{z}(\bm{r})=-2\left(\frac{DM}{Ja}\right)^{2}.
\end{align}
Since $\bar{\bm{f}}$ and $\bar{\bm{h}}$ are parallel, the system breaks the time-reversal symmetry and has the finite thermal Hall conductivity.

\subsubsection{Canted antiferromagnetic order with easy-plane anisotropy}
Next, we consider the case where $\bm{m}_{\mathrm{uni}}(\bm{r})=\bm{e}^{z}$ and $\bm{m}_{\mathrm{stg}}(\bm{r})=\bm{e}^{x}$ as shown in Fig.~\ref{fig:texture-gauge}(d), which can be realized when $\Lambda>0$, $\bm{B}=(0,0,B^{z})$, and $|\Lambda|$ is sufficiently larger than $|D|$. From Eqs.~(\ref{eq:Tx})-(\ref{eq:T=tsigma}), we obtain the nonzero elements of $\bm{t}_{\alpha}(\bm{r})$, $u(\bm{r})$, and $\bm{h}(\bm{r})$ as
\begin{align}
    t_{x}^{y}(\bm{r})&=-\frac{DM}{Ja},\\
    t_{y}^{z}(\bm{r})&=\frac{D}{Ja},\\
    u(\bm{r})&=8J\Lambda S^{2}+8JSg\mu_{\mathrm{B}}MB^{z},\\
    h^{x}(\bm{r})&=8J\Lambda S^{2},
\end{align}
The field strength is calculated from Eq.~(\ref{eq:f}) as
\begin{align}
    f^{x}(\bm{r})=-2M\left(\frac{D}{Ja}\right)^{2}.
\end{align}
Similar to the case of the canted antiferromagnetic order with the easy-axis anisotropy, the noncommutative structure of the SU(2) gauge field generates the field strength that is parallel to the effective Zeeman field, leading to the finite thermal Hall conductivity. The emergence of the finite thermal Hall response in thie case is already investigated by the linear spin-wave theory~\cite{kawano2019prb}.

\subsubsection{Spiral order with zero net magnetization}
We now turn to the case where the ground-state spin configuration is given by a spiral order as shown in Fig.~\ref{fig:texture-gauge}(e). Here, we assume that $|D/\Lambda|$ is sufficiently large and $\bm{B}=\bm{0}$, which leads to the spiral order with $M=0$. The similar calculations leads to
\begin{align}
    t_{x}^{z}(\bm{r})&=\left[\bm{A}_{\mathrm{stg}}(\bm{r})-\frac{D}{Ja}m_{\mathrm{stg}}^{y}(\bm{r})\right],\\
    t_{y}^{z}(\bm{r})&=\left[\bm{A}_{\mathrm{stg}}(\bm{r})+\frac{D}{Ja}m_{\mathrm{stg}}^{x}(\bm{r})\right],\\
    u(\bm{r})&=8J\Lambda S^{2}\left[(\bm{e}^{X}(\bm{r})\cdot\bm{e}^{z})^{2}-2(m_{\mathrm{stg}}^{z}(\bm{r}))^{2}\right],\\
    h^{x}(\bm{r})&=-8J\Lambda S^{2}(\bm{e}^{X}(\bm{r})\cdot\bm{e}^{z})^{2},
\end{align}
with $t_{\alpha}^{x}(\bm{r})=t_{\alpha}^{y}(\bm{r})=0$ ($\alpha=x,y$) and $h^{y}(\bm{r})=h^{z}(\bm{r})=0$. The nonzero element of the field strength comes from $\partial_{x}T_{y}(\bm{r})-\partial_{y}T_{x}(\bm{r})$ and is calculated as
\begin{align}
    f^{z}(\bm{r})&=\bm{m}_{\mathrm{stg}}(\bm{r})\cdot[\partial_{x}\bm{m}_{\mathrm{stg}}(\bm{r})\times\partial_{y}\bm{m}_{\mathrm{stg}}(\bm{r})]\nonumber\\
    &\quad+\frac{D}{Ja}[\partial_{x}m_{\mathrm{stg}}^{x}(\bm{r})+\partial_{y}m_{\mathrm{stg}}^{y}(\bm{r})].
    \label{eq:f_spiral_M=0}
\end{align}
The first term in Eq.~(\ref{eq:f_spiral_M=0}) is proportional to the skyrmion density, whose spatial average is zero in the spiral order. The spatial average of the second term in Eq.~(\ref{eq:f_spiral_M=0}) gives the surface integral and should vanish. Therefore, the spatially-averaged field strength is zero, $\bar{\bm{f}}=\bm{0}$, indicating the zero thermal Hall conductivity.

\subsubsection{Antiferromagnetic skyrmion crystals}
We finally consider the case where the spin configuration is given by the AFM-SkX as shown in Fig.~\ref{fig:texture-gauge}(f), which can be achieved in the system with finite $\Lambda$ and $\bm{B}$. For simplicity, we assume $M\simeq0$. Then we have nonzero elements of $\bm{t}_{\alpha}(\bm{r})$, $u(\bm{r})$ and $\bm{h}(\bm{r})$ as
\begin{align}
    t_{x}^{z}(\bm{r})&=\left[\bm{A}_{\mathrm{stg}}(\bm{r})-\frac{D}{Ja}m_{\mathrm{stg}}^{y}(\bm{r})\right]\sigma^{z},\\
    t_{y}^{z}(\bm{r})&=\left[\bm{A}_{\mathrm{stg}}(\bm{r})+\frac{D}{Ja}m_{\mathrm{stg}}^{x}(\bm{r})\right]\sigma^{z},\\
    u(\bm{r})&=8J\Lambda S^{2}\left[(\bm{e}^{X}(\bm{r})\cdot\bm{e}^{z})^{2}-2(m_{\mathrm{stg}}^{z}(\bm{r}))^{2}\right],\\
    h^{x}(\bm{r})&=-8J\Lambda S^{2}(\bm{e}^{X}(\bm{r})\cdot\bm{e}^{z})^{2},\\
    h^{z}(\bm{r})&=8JSg\mu_{\mathrm{B}}\bm{B}\cdot\bm{m}_{\mathrm{stg}}(\bm{r}).
\end{align}
The important difference is that, because of the topologically-nontrivial spin texture in real space, the spatially-averaged field strength takes a finite value
\begin{align}
    \bar{f}^{z}=4\pi\rho_{\mathrm{SkX}}\neq0,
\end{align}
where $\rho_{\mathrm{SkX}}$ is the skyrmion density
\begin{align}
    \rho_{\mathrm{SkX}}=\frac{1}{4\pi V}\int\mathrm{d}^{2}\bm{r}\ \bm{m}_{\mathrm{stg}}(\bm{r})\cdot[\partial_{x}\bm{m}_{\mathrm{stg}}(\bm{r})\times\partial_{y}\bm{m}_{\mathrm{stg}}(\bm{r})].
    \label{eq:skx}
\end{align}
Together with $\bar{h}^{z}=8JSg\mu_{\mathrm{B}}\bm{B}\cdot\bar{\bm{m}}_{\mathrm{stg}}$, the system satisfies the condition to break the time-reversal symmetry and we expect the finite thermal Hall response in the AFM-SkX phase.

\section{Conclusions}
\label{sec:conclusion}
In this work, we developed the field-theoretical framework to elucidate the condition to have the thermal Hall effect of magnons in square-lattice antiferromagnets. By constructing an effective field theory from the spin model, we demonstrated that magnons in two-sublattice antiferromagnets acquire an effective SU(2) gauge field that couples to their pseudospin degrees of freedom, which originate from the underlying two-sublattice structure. The field strength associated with the SU(2) gauge field acts as a pseudospin-dependent magnetic field, bending the propagation of magnons with opposite pseudospins in opposite directions. An effective Zeeman field generates an imbalance between pseudospin-up and pseudospin-down magnons, thereby enabling a finite thermal Hall response. We also applied our framework to four representative spin textures: easy-axis canted antiferromagnetic order, easy-plane canted antiferromagnetic order, spiral order, and AFM-SkXs. Then we clarified how the combination of the DM interaction, single-ion anisotropy, external magnetic field, and real-space spin texture breaks the effective time-reversal symmetry, which must be broken for a finite thermal Hall response to appear. Beyond the square-lattice antiferromagnets, our approach can be extended to a broader class of magnetic systems, including noncoplanar spin textures as well as multi-sublattice antiferromagnets. Exploring how differences in the structure of gauge fields affect transport phenomena may be one of the directions for future research.

\bibliography{biblio}

\end{document}